\documentclass[twocolumn,showpacs,preprintnumbers,amsmath,amssymb]{revtex4}
\usepackage{amsmath,amsfonts,latexsym,amssymb,graphics,epsfig,subfigure,color,makeidx}
\usepackage{xcolor}
\usepackage[utf8]{inputenc}
\usepackage{graphicx,hyperref}
\usepackage{bm}
\usepackage{color}
\usepackage{multirow}

\begin{document}
	
	\title{Emerging black hole shadow from collapsing boson star}
	\author{Yu-Peng Zhang\footnote{zhangyupeng@lzu.edu.cn},
		Shao-Wen Wei\footnote{weishw@lzu.edu.cn},
		Yu-Xiao Liu\footnote{liuyx@lzu.edu.cn, corresponding author}
	}
	\affiliation{
		Key Laboratory of Quantum Theory and Applications of MoE, Lanzhou Center for Theoretical Physics,\\
	 Key Laboratory of Theoretical Physics of Gansu Province, Gansu Provincial Research Center for Basic Disciplines of Quantum Physics, Lanzhou University, Lanzhou 730000, China\\
	    Institute of Theoretical Physics \& Research Center of Gravitation, School of Physical Science and Technology, Lanzhou University, Lanzhou 730000, China
	}
	
	\begin{abstract}
This work devotes to investigate the dynamical emergence of black hole shadow from gravitational lensing in dynamical spacetime by using the collapsing boson star. Two characterized scenarios are adopted with or without considering the time delay of light propagation. As the boson star evolves, new Einstein rings emerge from the lensing center, with their radius gradually increasing, and their number continues to grow infinitely before the light-ring forms. The shadow forms instantaneously at the moment the black hole appears when ignoring the time delay of light propagation. Considering the time delay for light propagation in dynamical spacetime, a more intricate process of the shadow formation is uncovered: it first appears as a minute dot in the lensing center, then gradually grows as the black hole grows, eventually expands the inner region of the light-ring. During the quasi-stable phases of boson star and black hole, the lensing and shadow structures from two scenarios are nearly identical and remain almost unchanged. Our results present the universal dynamic patterns of the lensing and shadow structures, and reveal the potential observed phenomena near the collapsing star and the event horizon of the newly formed black hole.

	\end{abstract}
	\maketitle


\section{Introduction}

The recently observations of gravitational waves \cite{Abbott2016a,LIGOScientific:2018jsj,LIGOScientific:2020kqk,LIGOScientific:2021psn} and black hole images \cite{eth2019,eth2022} provided the groundbreaking evidence for the existence of black holes in our Universe. They also opened up new frontiers in astrophysics, enabling us to study black holes and their properties in ways previously thought impossible. Impressively, the photons flying past the center black hole will be absorbed or bent by its extremely strong gravity. Accordingly, it leaves a dark zone, the black hole shadow, in the celestial sphere of the observer, and around which a series lensing rings are presented. Such observed signature offer a crucial tool to test the direct information about the size, shape, and the surrounding spacetime curvature of the static or stationary center black holes \cite{Luminet:1979nyg,Falcke:1999pj,Chen:2009eu,Herdeiro:2014goa,Cunha:2015yba,Cunha:2017wao,Gralla:2019xty,Herdeiro:2021lwl,Perlick:2021aok,Chen:2022scf,Kuang:2024ugn,Huang:2024gtu,Cunha:2018acu}.

Although there have been many studies on the lensings and shadows of static or stationary black holes, their dynamical features during the process of collapsing into a black hole from a compact star have not been well revealed. As a star collapses, light follows a path that transitions from gravitational lensing, where light is merely bent, to absorption by the newly-formed black hole, creating the black hole's shadow \cite{Vincent:2012kn}. This process provides a unique opportunity to test general relativity in strong gravitational fields, especially the dynamics of the event horizon's formation \cite{Choptuik:1992jv,Gundlach:2007gc}.

Now, it is extensively known that some Boson stars, one kind hypothetical exotic compact objects are constructed by scalar fields, are dynamical unstable \cite{Seidel:1990jh,Balakrishna:2006ru,Sanchis-Gual:2019ljs,Siemonsen:2020hcg,Dmitriev:2021utv,Liebling:2012fv,Sanchis-Gual:2021phr,Cunha:2022gde}. During their evolutions, they will encounter collapses from a loosely bound matter distribution to a highly compact object and finally form black holes \cite{Seidel:1990jh,Balakrishna:2006ru,Cunha:2017wao,Sanchis-Gual:2019ljs,Siemonsen:2020hcg,Dmitriev:2021utv,Liebling:2012fv,Sanchis-Gual:2021phr,Cunha:2022gde,Zhang:2023qag}. Therefore, in this Letter, we shall adopt such excellent model to test the complete dynamical evolution of gravitational lensing and shadow formation during the oscillations and collapse of a boson star into a black hole. The universal dynamical features of gravitational lensing and black hole shadow in such process are expected to be uncovered.

\section{The model}

As the first step for uncovering the dynamical evolution of the spherical boson star, let us consider the following action \cite{Sanchis-Gual:2021phr}
	\begin{equation}
	S = \int d^4x \sqrt{-g}
	\bigg(\frac{R}{16\pi}-\frac{1}{2}\partial_\mu\Phi \partial^\mu\Phi^*-V(|\Phi|^2)\bigg),
	\label{action}
	\end{equation}
where the potential term reads
	\begin{equation}
	V(|\Phi|^2)=\frac{1}{2}\mu^2|\Phi|^2+\frac{\lambda}{4}|\Phi|^4.
	\end{equation}
The parameters $\mu$ and $\lambda$ are the mass and self-interacting parameters, respectively. We set $\mu=1$ and the units with $G=c=1$. The introduction of the self-interaction $\frac{\lambda}{4}|\Phi|^4$ can stabilize the excited spherical boson star with the large value of $\lambda$ and hence one can obtain the stable and unstable boson stars by choosing the proper values of $\lambda$ \cite{Sanchis-Gual:2021phr}.

The solution of spherical boson stars is initially stationary and the scalar field satisfies $\Phi=\phi(r)\exp(i \omega t)$, the parameter $\omega$ is the oscillation frequency. Given the stationary solutions of boson stars as the initial data, we perform the simulation of the boson star and obtain the dynamical spacetime that including the oscillation of boson star and formation of black hole. The spacetime evolution is realized by using our spherical numerical relativity code \cite{Zhang:2023qag} in terms of the Baumgarte-Shapiro-Shibata-Nakamura formalism in spherical coordinates~\cite{Montero:2012yr}.

To show the lensing and shadow caused by the boson star or black hole, we divide the celestial sphere into four quadrants painted with four different colors, we use a grid of constant longitude and latitude lines separated by $\pi/16$. As the setup of Ref. \cite{Bohn:2014xxa}, we place the observer inside the celestial sphere at an off-centered position and it has a viewing angle $\pi/3$, the distance between the observer and the center of the boson star or black hole is $r_0=30/\mu$. We span the viewing angle of the observer with $800\times800$ pixels. In what follows, we will consider two scenarios for calculating the gravitational lensing and black hole shadows.

In Ref. \cite{Cunha:2015yba}, the authors have obtained the shadow structures of different rotating hairy black holes that can continuously transition within the parameter space of hairy black holes. For the entire process of a boson star collapsing into a black hole, the corresponding spacetime slices at every moment can also provide such a similar continuous transition. Therefore, the first scenario treats the every spacetime slices as fixed background and directly computes the trajectories of null geodesics. Although the resulting light trajectories do not capture the dynamic changes of spacetime, this assumption is helpful for us to analyze the variations in lensing and shadows across different compactness states throughout the entire process, from gradual changes in compactness to the formation of a black hole.

Considering the fact that the spacetime evolves over time as the light propagates, in the second scenario, we consider the time delay of light and calculate the geodesic integration backwards in time in parallel with the spacetime integration by interpolating metric from the numerical grid \cite{Zhang:2023qag}. However, for the collapse of a boson star into a black hole, we cannot simply obtain the pre-collapse boson star by time-reversing the final black hole state. To overcome this difficulty, we obtain the background spacetime at different time intervals during the dynamic process of boson star collapse. Then we read the background spacetime data and perform the time-reversed geodesic evolution within the dynamic spacetime. Combining our spherical numerical relativity code \cite{Zhang:2023qag}, we have developed a code for computing the gravitational lensing and black hole shadow in spherical dynamical spacetime. The corresponding numerical integration of the geodesics are realized by using the the $3+1$ form of the geodesic \cite{Vincent:2012kn,Bohn:2014xxa}.

Due to the geometry of the dynamical background is evolving with time, we should solve the initial data of the null geodesics at different instances. It is impossible that we compute all the lensing and shadow at every instances. We introduce a parameter $\Delta t_1=0.25M_0$ that specifies the time interval for calculating the lensing or shadow in the dynamic spacetime. Another timescale is involved in solving the motion of light rays: the duration each light ray must propagate from the observer to produce the desired lensing and shadow results for each calculation. Ideally, each light ray would travel for an infinite amount of time to determine its final position on the celestial sphere. However, in practical computations, infinite time is not feasible, we can compute the motion of light rays over a finite duration $\Delta t_2$. We tested different time scales $\Delta t_2$ to validate the corresponding lensing and shadow results. We found that when the time scale is set to $\Delta t_2=200M_0$, the lensing and shadow results stabilize and no longer show significant changes.

In the first scenario, we only need to calculate null geodesics on different spacetime slices. However, in the second scenario, the number of geodesics to be computed increases with the system's evolution time. Specifically, for each additional interval $\Delta t_1$, the number of geodesics doubles. When the spacetime evolution time exceeds the interval $\Delta t_2$ required to compute a single complete geodesic, the number of geodesics computed simultaneously reaches its maximum as follows
\begin{equation}
N=800\times 800\times\frac{\Delta t_1}{\Delta t_2}=800^3=512000000.
\end{equation}
We have utilized OpenMP parallelization to accelerate the geodesic calculations.

\section{Dynamical lensing and shadows}

In order to clearly show the complete dynamical evolution monitored by the variation of the minimum of the lapse function over time, we simulate the spherical excited boson star with $\omega=0.88\mu$ and $\lambda=0$ without loss of generality.

\begin{figure}[!htbp]
	\includegraphics[width=\linewidth]{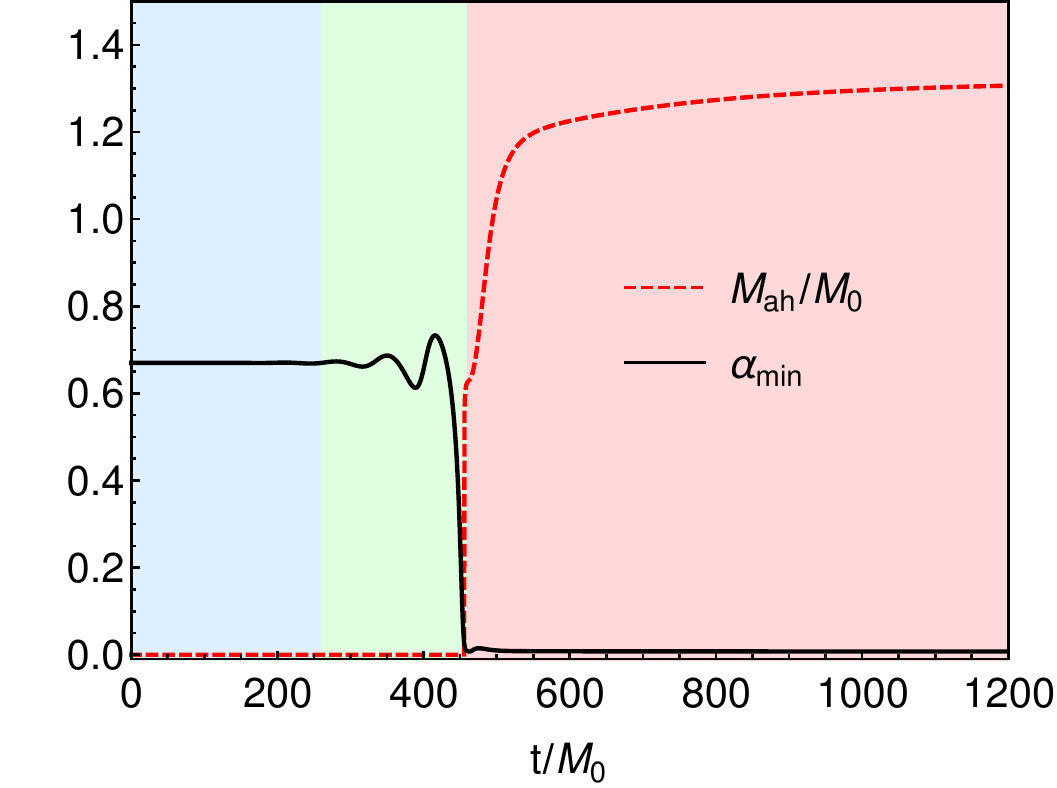}
	\caption{Evolution of the minimum of the lapse function $\alpha_{\text{min}}$ and the irreducible mass $M_{\text{ah}}$ of the newly formed black hole, where we define a mass parameter $M_0$ satisfies $M_0 \mu=1$. The light blue, light green, and light red regions describe the \emph{quasi-equilibrium}, \emph{oscillation-collapse}, and \emph{accretion-growth} stages respectively.}
	\label{p_m_phi_ah}
\end{figure}

The evolutionary characteristics of the minimum of the lapse function $\alpha_{\text{min}}$ and the irreducible mass $M_{\text{ah}}$ of the apparent horizon are exhibited in Fig. \ref{p_m_phi_ah}. This is highly consistent with the previous general results outlined in Refs. \cite{Zhang:2023qag,Sanchis-Gual:2021phr}. After detailed examination, this universal process can be divided into three stages: \emph{quasi-equilibrium}, \emph{oscillation-collapse}, and \emph{accretion-growth}.

In the quasi-equilibrium stage (0, 260$M_0$), the boson star and its surrounding scalar field keeps a fine balance. After the time $t=260M_0$, the system approaches to the oscillation-collapse stage. Such balance is gradually being disrupted, where the minimum lapse function shown in Fig. \ref{p_m_phi_ah} exhibits an oscillatory behavior. The oscillation amplitude gradually increases, and when it reaches a critical threshold, it decreases to zero in a very short time, indicating that the boson star suddenly collapses and forms a black hole. Once the center black hole is formed, the structure of spacetime encounters a huge change, and therefore the system turns to the accretion-growth stage. As the scalar field flows into the event horizon, the black hole gradually becomes larger, which also further enhances its attraction. Eventually, almost all scalar fields are absorbed into the black hole, turning the entire spacetime into a Schwarzschild spacetime. Moreover, through the evolution of the irreducible mass $M_{\text{ah}}$ marked with the red dashed line in Fig. \ref{p_m_phi_ah}, it is obvious that the star gradually becomes more compact, starting from an initially diffuse structure in the whole stage.

After fully understanding this evolution, we will study the dynamic variations of the gravitational lensing and shadow. There are two different scenarios for one to study. The first scenario treats each evolutionary state as fixed spacetime background. Then the lensing and shadow can be explored. The second scenario counts the time delay of photons traveling from the black hole to the observer. Considering the propagation time of light and the dynamic changes of spacetime, we cannot provide the gravitational lensing structure at the initial time point $t=0M_0$. Therefore, we begin by presenting the corresponding gravitational lensing at the time point $t=200M_0$. The patterns of the lensing and shadow shall reveal the nature of the spacetime when the system continuously evolved from a diffuse distribution to a black hole.

For spherically symmetric boson stars, the bosonic field is primarily concentrated near the center of the system. Even in the oscillation and collapse stages, the essential change also occurs near its center. So in the regions far from the star's center, it resembles a Schwarzschild spacetime, and thus the outer gravitational lensing structure almost remains unchanged. This point shall be supported by our following numerical results. Moreover, early study shows that the time-dependent effective potential is feasible in monitoring the existence of light-ring in the dynamical spacetime \cite{Cunha:2022gde}, so we will also adopt such technique.

\begin{figure}[!htbp]
	\includegraphics[width=\linewidth]{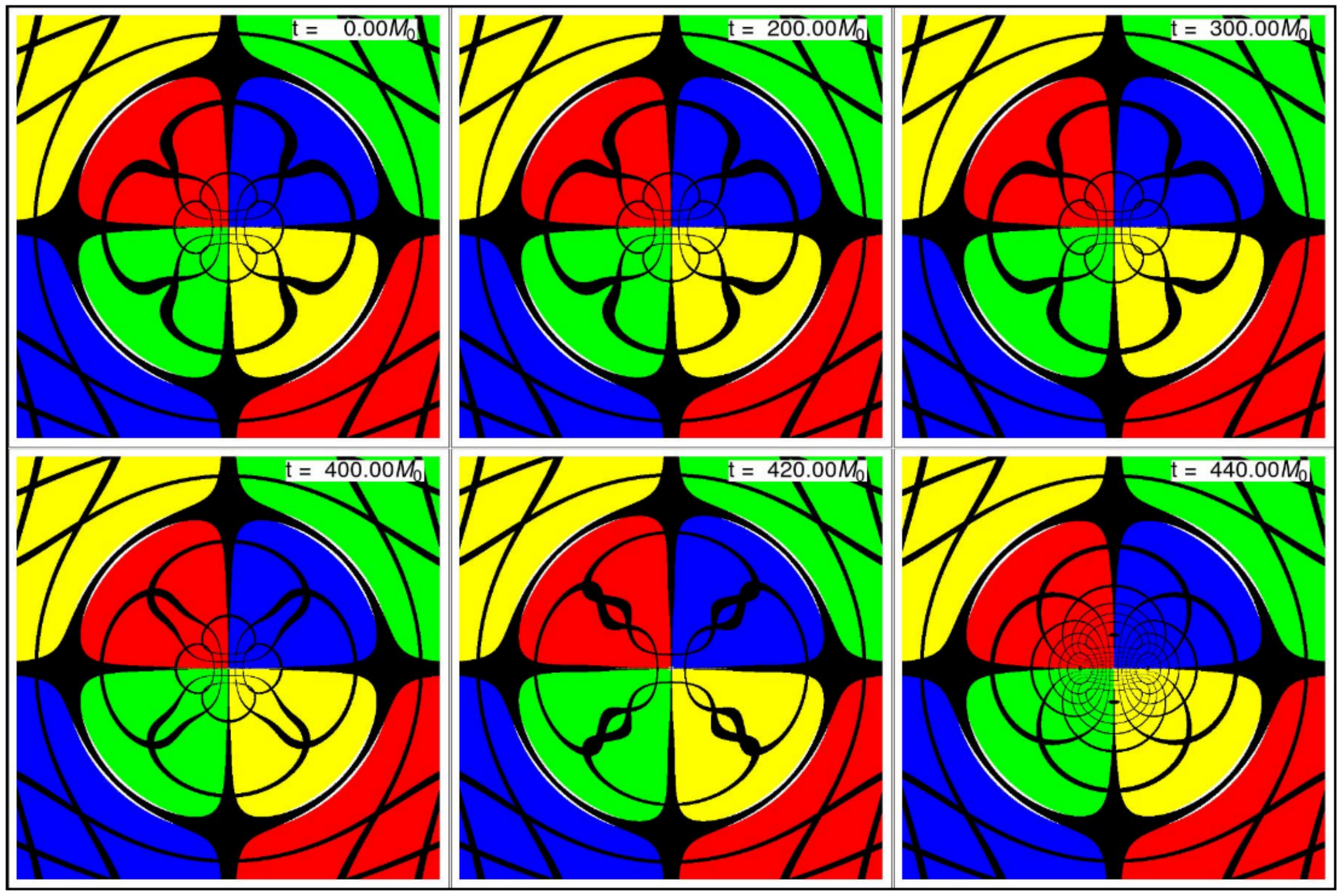}
	\caption{Oscillation of the gravitational lensing in the dynamical background for the first scenario.}
	\label{earlystage_lensing}
    \end{figure}

In the quasi-equilibrium stage, the boson star undertakes no obvious change via interacting with the scalar field, which results in a lensing structure, a large Einstein ring, that remained largely unchanged. As time progresses, the unstable modes accumulate and the boson star begins to oscillate. Consequently, during this oscillation-collapse stage, the structure of the gravitational lensing must change in response to the evolving background. As the oscillation amplitude of the boson star gradually increases, the grid line structure of the corresponding gravitational lensing begins to vary. The results are presented in Fig. \ref{earlystage_lensing}.

The lensing structures shall experience several nontrival periods before the formation of the black hole. The first distinctive feature is that, near $t=442.75M_0$, a new small Einstein ring emerges at the center. This is because that the scalar field condenses towards the center such that the light has a chance to surround the center for one loop. Another result is that the grid lines within the lensing structure begin to oscillate, and during this process, the density of the grid lines gradually increases. After the first new Einstein ring, the second, third, ..., rings gradually emerge from the center. Further with the evolution of the boson star, the number and radius of these newly formed rings gradually increases.

The next distinctive feature takes place when the light-rings appear at $t=452.00M_0$, which indicates that the center star has a high enough intense. An infinite number of Einstein rings can appear near the inner and outer regions of the light-rings in the corresponding lensing structure. From the effective potential, we confirm that a pair of the light-rings are produced simultaneously. One is radial stable, while other one is unstable. Within the stage from the formation of the light-ring to just before the black hole formation, the density of the boson star continues to increase. However, we further find that the radius of the newly formed unstable light-ring will no longer grows and remains almost unchanged. After a short time, the star encounter a sudden collapse and a new tiny black hole of almost zero mass is formed. Due to the topological charge of the light-rings changes from 0 to -1 \cite{Cunha:2020azh}, the pattern of the lensing and shadow structures in this period is expected to greatly change.

\begin{figure}[!htbp]
	\includegraphics[width=\linewidth]{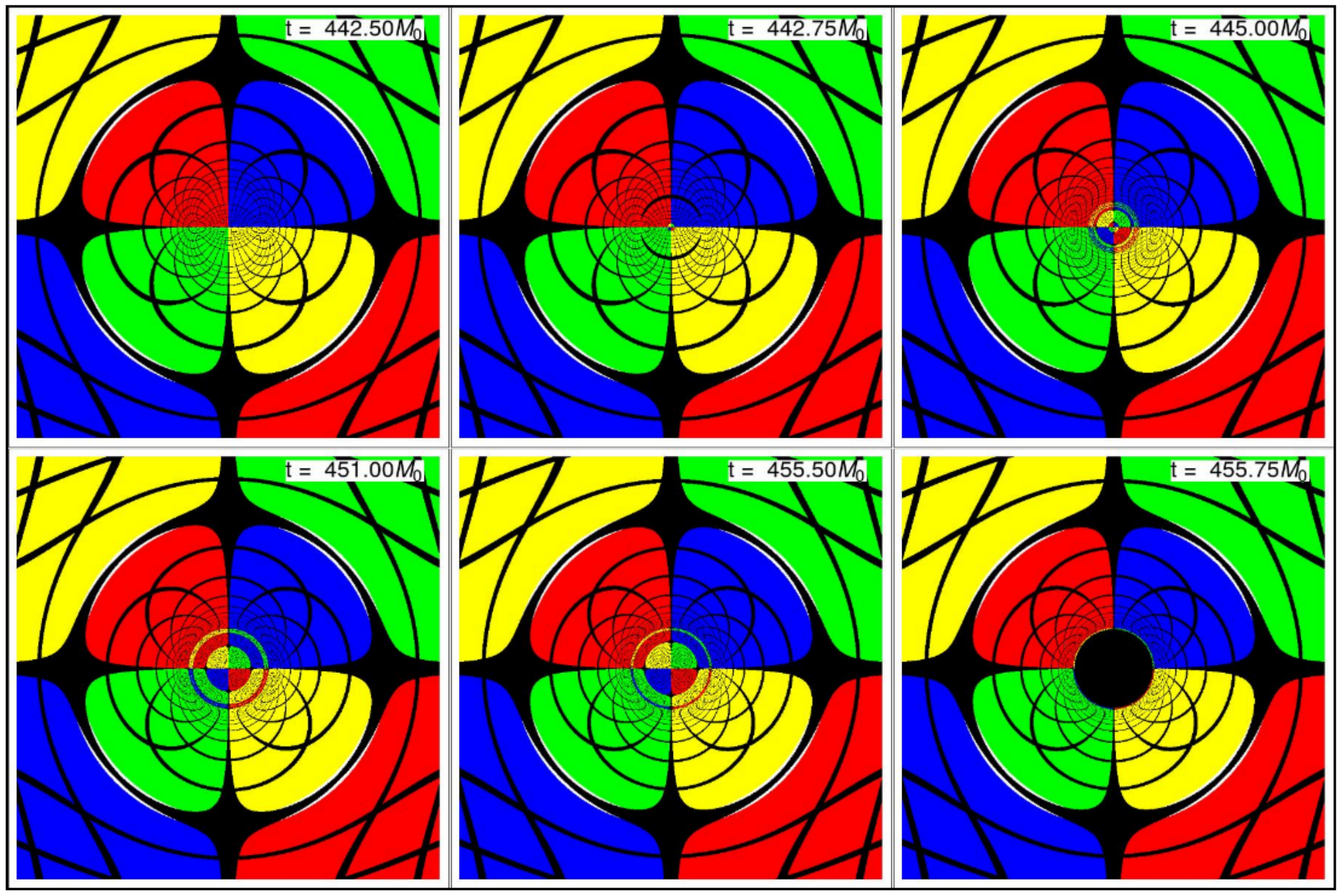}
	\caption{Transition of gravitational lensing and formation of shadow with first scenario.}
	\label{p_stationary_accretion_shadow}
    \end{figure}

Following the first scenario, each evolution state is treated as a fixed background. This can also be alternatively understood as no time delay in this case.  So once the black hole is formed, the event horizon appears. Simultaneously, the inner stable light-ring hides behind the horizon. Thus, only the unstable one presents outside black hole, which leads to a finite size shadow. As a result, accompanying the birth of a new black hole, a finite-sized black hole shadow is also instantaneously generated. The corresponding results are given in Fig. \ref{p_stationary_accretion_shadow}.

\begin{figure}[!htbp]
	\includegraphics[width=\linewidth]{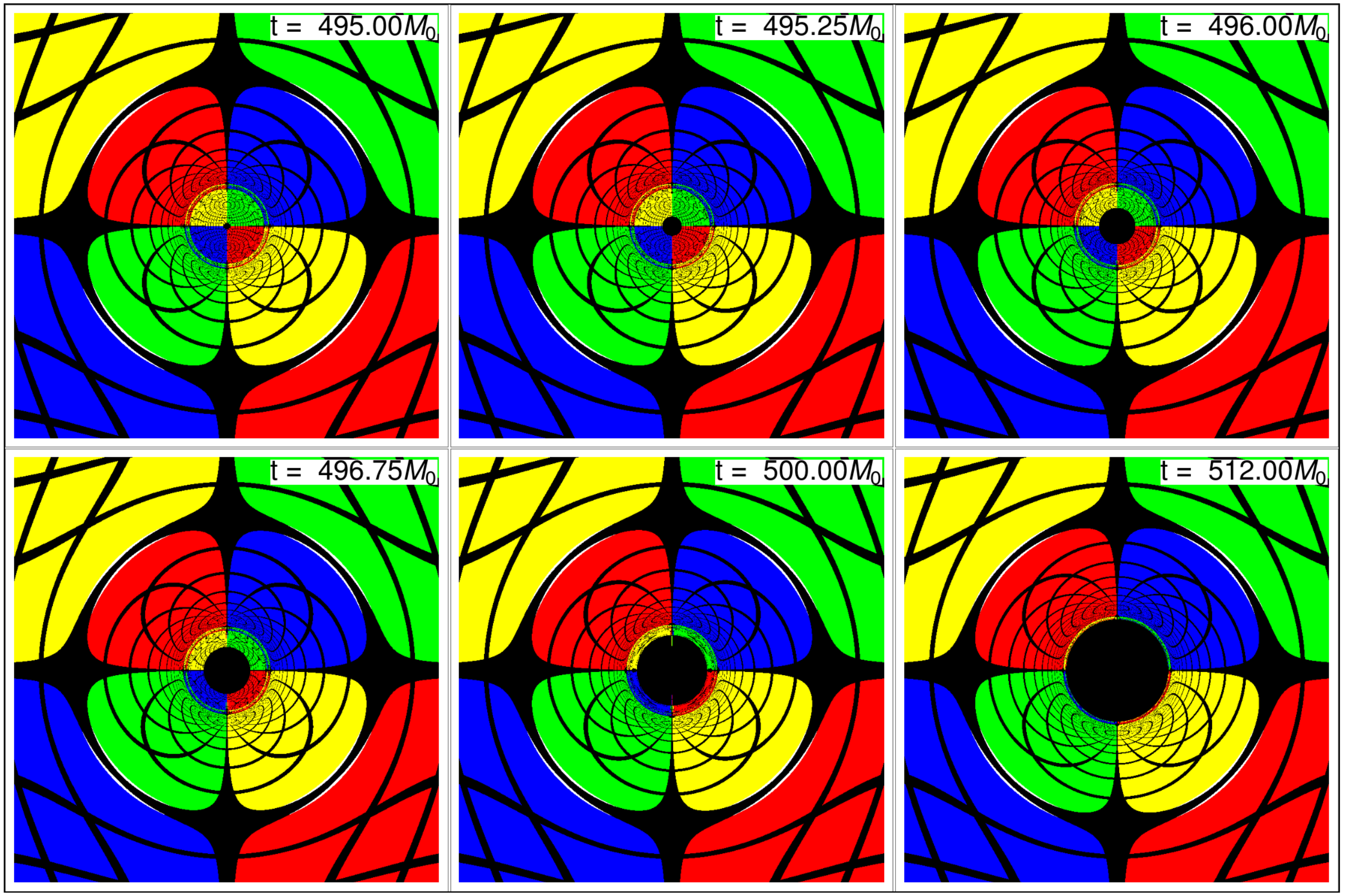}
	\caption{Transition of gravitational lensing and formation of shadow with second scenario.}
	\label{shadowgorw_dynamical_lensing}
\end{figure}

We turn to the second scenario: what does a real distant observer see? Considering the fact that the spacetime is evolving when the light propagates in a dynamical spacetime. Therefore we need to take account of the time delay of the light and the dynamic changes in the spacetime background during the propagation of light. Fur the purpose, we will track the trajectories of light rays emitted from fixed spatial coordinates backwards in time at different moments. We provide the results in Fig. \ref{shadowgorw_dynamical_lensing}. Obviously, even when the black hole is formed at $t=455.75M_0$, the observer will not see a different lensing structure until the short time delay $\Delta t=39.5M_0$ is reached. Significant difference is that the black hole shadow not appears simultaneously, but grows from a minute dot at its center and gradually expands.

\begin{figure}[!htbp]
	\includegraphics[width=\linewidth]{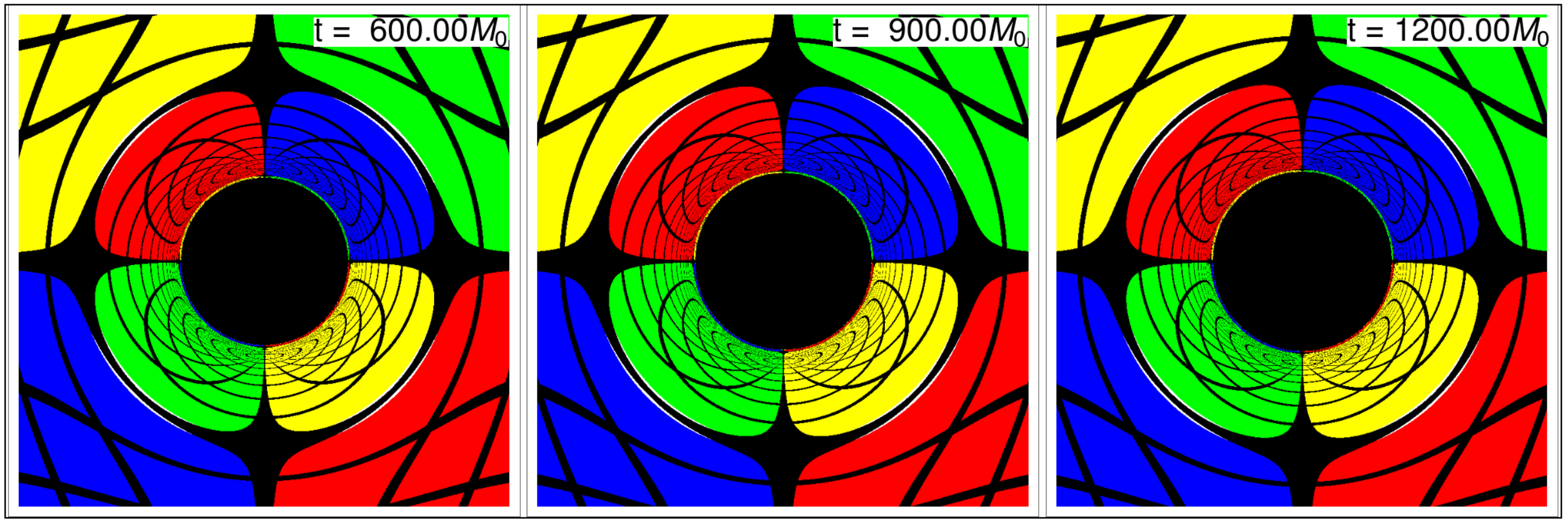}
	\caption{Growth of shadows at late time for two scenarios.}
	\label{shadowgorw_steady}
\end{figure}

After the formation of the black hole shadow, the black hole continually slowly accretes the background scalar field and takes a steady growth behavior. No significant changes of the lensing structure undergoes both for two scenarios. The feature is that the black hole shadow expands until all the scalar field is accreted. Finally a Schwarzschild black hole is left. We provide the results in Fig. \ref{shadowgorw_steady}.

\begin{figure}[!htbp]
	\includegraphics[width=\linewidth]{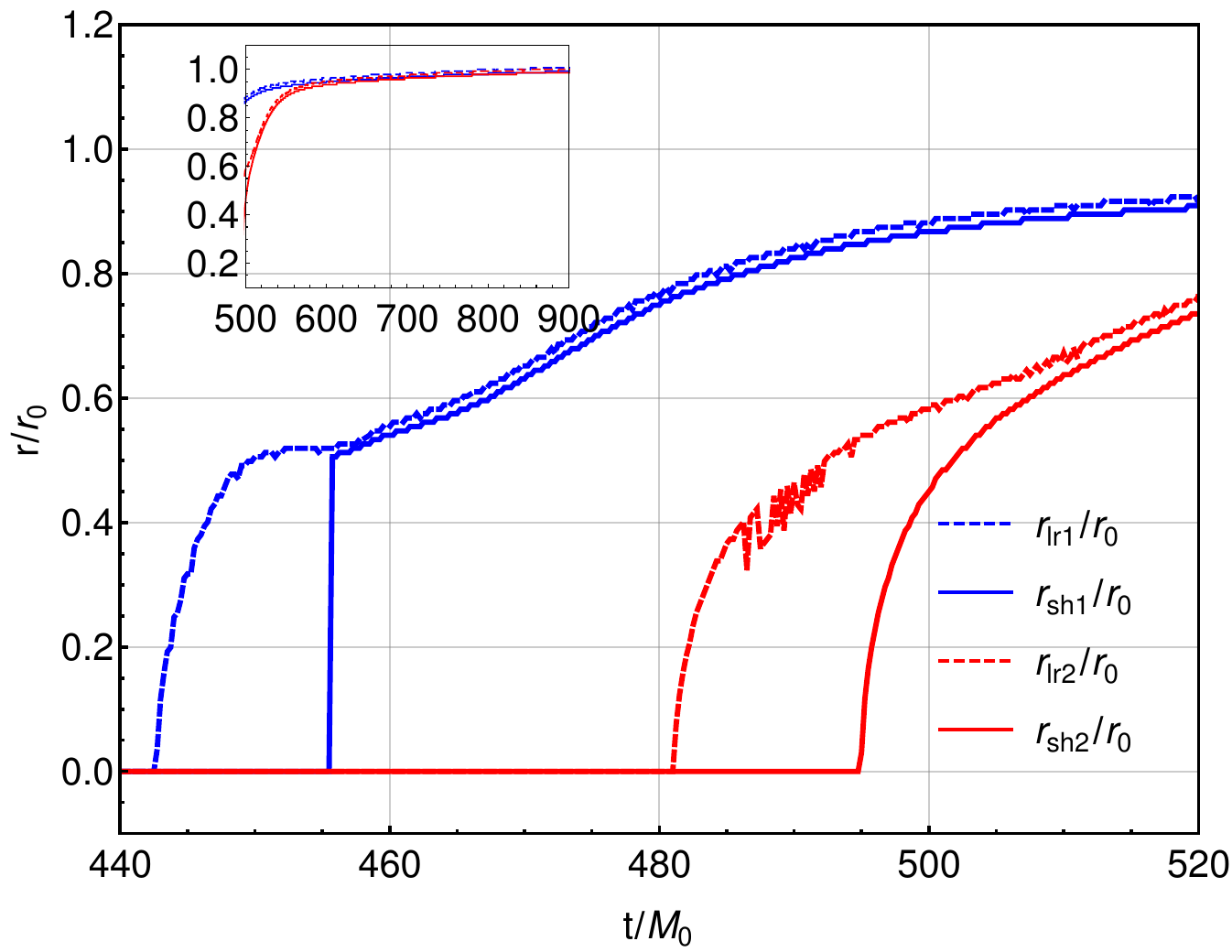}
	\caption{Time dependent of the radii for the first inner Einstein rings and black hole shadow in the dynamical background, where dashed lines stand for the radii of the first inner Einstein rings, solid lines stand for the radii of black shadows, the blue and red are related to the first and second scenarios, respectively. The inset figure is a zoom out of the figure.}
	\label{shadowgorw_radius}
\end{figure}

It can be seen that the outer weak lensing ring almost keeps the same size, this confirms the result that the change of the spacetime structure occurs near the collapsing center. The outer far region can be well described by a Schwarzschild black hole. In order to quantitatively exhibit the lensing and shadow structure, we provide the results that how the radii of light-ring and shadow evolve with time in Fig. \ref{shadowgorw_radius}.  The radii of first inner Einstein rings are described by the dashed blue ($r_\text{ps1}$) and red ($r_\text{ps2}$) lines for the first and second scenarios, respectively. Obviously, comparing the first scenario, the appearance and increase of the ring for the second scenario follow it and has a time delay about $\Delta t=39.5M_0$ measuring the light traveling from the black hole to the observer. During the accretion-growth stage, the spacetime background tends to the final stationary state, their size tends to the same as expected for the reason that the time delay has no obvious influence on the spacetime structure.

Moreover, we also show the radius of the black hole shadow with the evolutionary time in Fig. \ref{shadowgorw_radius} both for two scenarios, they are described by the solid blue ($r_\text{sh1}$) and red ($r_\text{sh2}$) lines for the first and second scenarios, respectively. In the quasi-equilibrium stage, no shadow is observed for both the scenarios, or one can say the radius of shadow is 0. While in the oscillation-collapse stage, once the black hole is formed, the shadow radius for the first scenario gets a sudden increase from 0 to $0.50545 r_0$, where $r_0$ is the radius of the shadow of the pure Schwarzchild black hole at late time. This exactly reflects the formation of the event horizon, a one way boundary for matter and lights. Quite differently, the shadow size for the second scenario by accounting the time delay increases from zero at $t=495M_0$ and expands following the black hole expansion. At the final stage, the size of the shadows gradually increase and eventually tend to $r_0$ for both the two scenarios, see the results from the inset figure in Fig. \ref{shadowgorw_radius}.

Our results indicates that for distant observers, the time delay would eliminate the sudden changes in the lensing and shadow structures during the collapse of the boson star to form a black hole.  It also reveals the actual formation process of a black hole shadow from the collapsing boson star.

\section{Summary}\label{Conclusion}

In this work, we first studied the dynamical evolution of the spherical boson star structured by self-interacting scalar field. Based on the distinct characteristics of the evolutionary process, we have divided it into three stages, quasi-equilibrium, oscillation-collapse, and accretion-growth stages. Then we investigated the lensing and shadow structures during these three stages, and their features are revealed for a distant observer during the evolutionary process of the boson star.

At the first stage, the boson star undergoes a quasi-equilibrium period. The lensing pattern keeps almost the same. However, the unstable modes accumulate and the boson star begins to oscillate at the second stage. A series new Einstein rings emerge from the center and enlarge with the time. In particular, at the last of this stage, the boson star collapses and a black hole forms. The lensing pattern gets an obvious change. For a distant observer by accounting the time delay, he will observe that the emergence of the black hole's shadow, initially appearing as a minute dot at its center and gradually expanding as the black hole grows regardless of the significant changes in the spacetime structure. At the last stage, the lensing rings and shadow experience steady growth until the black hole absorbs all the scalar field and turns to a Schwarzschild black hole.

More importantly, we observed that the shadow emerges from a minute dot or a finite size with or without considering the time delay of the propagation of light in dynamical spacetime. We provided the results for describing the universal dynamic patterns of the lensing and shadow structures in the background of collapsing boson star for a distant observer. These features would help us peeking into the strong gravitational regions near the collapsing star at final stage and the event horizon of the newly formed black hole.

\acknowledgments
	
This work was supported in part by the National Key Research and Development Program of China (Grant No. 2021YFC2203003), the National Natural Science Foundation of China (Grants No. 12105126, 12475056, and 12247101), the 111 Project under (Grant No. B20063), the Gansu Province's Top Leading Talent Support Plane.

\end{document}